\documentclass[aps,prb,twocolumn,showpacs,superscriptaddress,floatfix,
amsmath,amssymb,
]{revtex4-2}

\usepackage{graphicx}
\usepackage{dcolumn}
\usepackage{bm}
\usepackage{hyperref}
\usepackage[mathlines]{lineno}
\usepackage[american]{babel}

\begin{document}

\title{First-principles study of strain behavior in iron-based fluorides of tungsten bronze type as cathode materials for alkali ion batteries}

\author{A. F. Baumann}
 \email{aljoscha.baumann@fmf.uni-freiburg.de}
 \affiliation{University of Freiburg, Freiburg Materials Research Center (FMF), Stefan-Meier-Straße 21, 79104 Freiburg, Germany}
 
\author{D. Mutter}
 \affiliation{Fraunhofer IWM, Wöhlerstraße 11, 79108 Freiburg, Germany}

\author{D. F. Urban}
\affiliation{University of Freiburg, Freiburg Materials Research Center (FMF), Stefan-Meier-Straße 21, 79104 Freiburg, Germany}
\affiliation{Fraunhofer IWM, Wöhlerstraße 11, 79108 Freiburg, Germany}

\author{C. Elsässer}
\affiliation{University of Freiburg, Freiburg Materials Research Center (FMF), Stefan-Meier-Straße 21, 79104 Freiburg, Germany}
\affiliation{Fraunhofer IWM, Wöhlerstraße 11, 79108 Freiburg, Germany}

\begin{abstract}
Mechanical stresses and strains in the microstructure of cathode materials evolving during charge/discharge cycles can reduce the long-term stability of intercalation-type alkali-metal-ion batteries. In this context, crystalline compounds exhibiting \textit{zero-strain} (ZS) behavior are of particular interest. Near-ZS sodiation was experimentally measured in the \textit{tetragonal tungsten bronze} (TTB) type compound Na$_x$FeF$\mathrm{_3}$. Using a \textit{first-principles} method based on density functional theory, we investigate the potential of iron-based fluoride compounds with tungsten bronze (TB) structures as ZS cathode materials. Simulations were conducted to study the intercalation of the alkali metal ions Li$\mathrm{^+}$, Na$\mathrm{^+}$, and K$\mathrm{^+}$ into the TTB and two related TB structures of the cubic perovskite (PTB) and hexagonal (HTB) types. We describe compensating local volume effects that can explain the experimentally measured low volume change of Na$_x$FeF$\mathrm{_3}$. We discuss the structural and chemical prerequisites of the host lattice for ZS insertion mechanism for alkali ions in TB structures and present a qualitative descriptor to predict the local volume change, that provides a way for faster screening and discovery of novel ZS battery materials.
\end{abstract}

\maketitle


\section{Introduction}

The active cathode materials used in alkali ion batteries can undergo mechanical degradation during the charge and discharge cycles \cite{Edge_2021_degradation,Pender_2020_degradation,Han_2019_degradation}. This degradation is caused by phase transformations or irreversible changes in lattice parameters due to the intercalation and deintercalation of alkali ions. For example, structural transformations have been identified as the major cause of performance degradation in graphite/LiCoO$\mathrm{_2}$ lithium-ion batteries \cite{Wang_2007}. \textit{Zero-strain} (ZS) materials, characterized by their minimal volume changes during charge and discharge cycles, are considered as promising candidates for achieving mechanically stable cycling. This is particularly important in the development of all-solid-state batteries, where the stability of the interfaces between the active electrode and the solid electrolyte particles is of paramount importance \cite{koerver_2018,Wang_2023}.

Li$\mathrm{_4}$Ti$\mathrm{_5}$O$\mathrm{_{12}}$ (LTO) is a well-known ZS anode material, which has been thoroughly investigated \cite{Zaghib_1998_LTO_ZS,Ohzuku_1995_LTO_ZS, Ziebarth_2014_LTO}. Cathode materials for which ZS behavior has been measured experimentally include: LiCoMnO$\mathrm{_4}$ \cite{Ariyoshi_2018_ZS_LCOMNO4}, Li$\mathrm{_2}$Ni$\mathrm{_{0.2}}$Co$\mathrm{_{1.8}}$O$\mathrm{_4}$ \cite{Ariyoshi_2019_ZS_LiNiCoO4}, and Li$_x$CaFeF$\mathrm{_6}$ \cite{de_biasi_licafef_2017}. To understand the ZS behavior, atomistic simulations can help to reveal the mechanisms responsible for the small volume changes. The ZS behavior in LTO was explained by the compensation of changing O-Ti-O bond angles and Li-O bond lengths during (dis)charging \cite{Tian_2020_LTO_ZS}. In the disordered rock-salt-type Li-excess compounds Li$\mathrm{_{1.3}}$V$\mathrm{_{0.4}}$Nb$\mathrm{_{0. 3}}$O$\mathrm{_2}$ and Li$\mathrm{_{1.25}}$V$\mathrm{_{0.55}}$Nb$\mathrm{_{0.2}}$O$\mathrm{_{1.9}}$F$\mathrm{_{0. 1}}$, the ZS behavior is described by effects including the presence of transition metal redox centers and a migration of Li from octahedral to tetrahedral sites \cite{Zhao_2022}. The local volume changes in the colquiriite compounds Li$_x$CaMF$\mathrm{_6}$, with $M$ denoting \textit{3d} transition metals (TM), were described in Ref. \cite{Baumann_Col}, where a compensation mechanism of expanding MnF$\mathrm{_6}$ octahedra and shrinking LiF$\mathrm{_6}$ octahedra was identified to be responsible for a small volume change of the unit cell of Li$_x$CaMnF$\mathrm{_6}$.

Near-ZS behavior was also reported for Na$_x$FeF$\mathrm{_6}$ in the tetragonal tungsten bronze structure (TTB) \cite{Han_2016_ZS_TB}. Tungsten bronzes (TB) are named after compounds of the type $A_x$WO$\mathrm{_3}$ and generally refer to compounds whose basic structure consists of a transition metal element (originally tungsten, in this paper iron) octahedrally surrounded by anions (originally oxygen, here fluorine). Another metal element, $A$, occupies sites in the structure with concentration $x$ \cite{Dickens_1968_TB_general}. Compounds of this structure family have in common that the octahedra surrounding the transition metal ion are corner-sharing and thus form a base structure which, depending on the arrangement, exhibits different intercalation sites. 

The use of TB type structures as active electrode materials has already been reported: The compound KFeF$\mathrm{_3}$ in the perovskite tungsten bronze structure (PTB), in which the potassium cations act as channel fillers during synthesis, was suggested as possible Li and Na ion cathode materials \cite{Martin_2019_PTB_K_cathode,Cao_2017_PTB_cathode}. The hexagonal tungsten bronze type structure (HTB) H$\mathrm{_{0.25}}$Cs$\mathrm{_{0.25}}$Nb$\mathrm{_{2.5}}$W$\mathrm{_{2.5}}$O$\mathrm{_{14}}$ has been proposed as a possible Li ion anode material, showing excellent capacities at high rates \cite{Calvez_2023_HTB_anode}. This good performance was attributed to the one-dimensional channels of the structure, which provide good diffusion properties. Similar arguments were made for  \nolinebreak{HTB FeF$\mathrm{_3}$-0.33H$\mathrm{_2}$O} \cite{Li_2011_Li_cathode}.

In this work we use atomistic simulations, based on density functional theory (DFT), to unravel the local mechanisms in Na$_x$FeF$\mathrm{_3}$, which are responsible for the near ZS behavior reported in literature. Subsequently, we extend the investigation to the iron-based fluorides of PTB and HTB type structures. In all three structures we systematically investigate the intercalation of the alkali metal ions $A$=Li$\mathrm{^+}$, Na$\mathrm{^+}$ and K$\mathrm{^+}$. For the TTB structure it is known that orthorhombic distortions can occur depending on the temperature and, especially relevant for this work, on the concentration of the $A$ cations \cite{Whittle_2018_TB_dist,Mezzadri_2008_TB_dist,Reisinger_2011_TB_dist}. Throughout this paper, the abbreviations PTB, HTB and TTB are used for crystal structures with a topology corresponding to the parent types, in which the structures were setup.

The paper is organized as follows: In Sec.\ \ref{sec:methods_structure}, the investigated TB-type structures are introduced, followed by a description of the computational details in Sec.\ \ref{sec:methods_compdetails}. In Sec.\ \ref{sec:results} we present the results, starting with the ground state structures in relation to magnetic arrangements of the iron ions (\ref{res:results_magnetic}). The general procedure for the determination of global unit-cell and local nieghbor-polyhedron volume changes at various concentrations is explained in Sec.\ \ref{sec:result_naxfef3} for the example of TTB Na$_x$FeF$\mathrm{_3}$ and the results are compared to experimental data. We elucidate the compensation mechanism of different local volume changes leading to a small global volume change. Results are then presented for the three structures in which the different $A$ cations are incorporated (\ref{sec:result_systematic}). In Sec.\ \ref{sec:discussion} we first discuss the suitability of TB type structures as ZS cathode materials (\ref{disc:TB_as_cathode}). Then the precondition of the structures for a suitable compensation mechanism is discussed and we derive a qualitative design criterion (\ref{disc:quali_ZS}). Our conclusions are given in Sec. \ref{sec:conclusion}. 


\section{Theoretical approach} \label{sec:methods}

\subsection{TB-type structures}\label{sec:methods_structure}

\begin{figure}
    \centering
    \includegraphics[width=1\columnwidth]{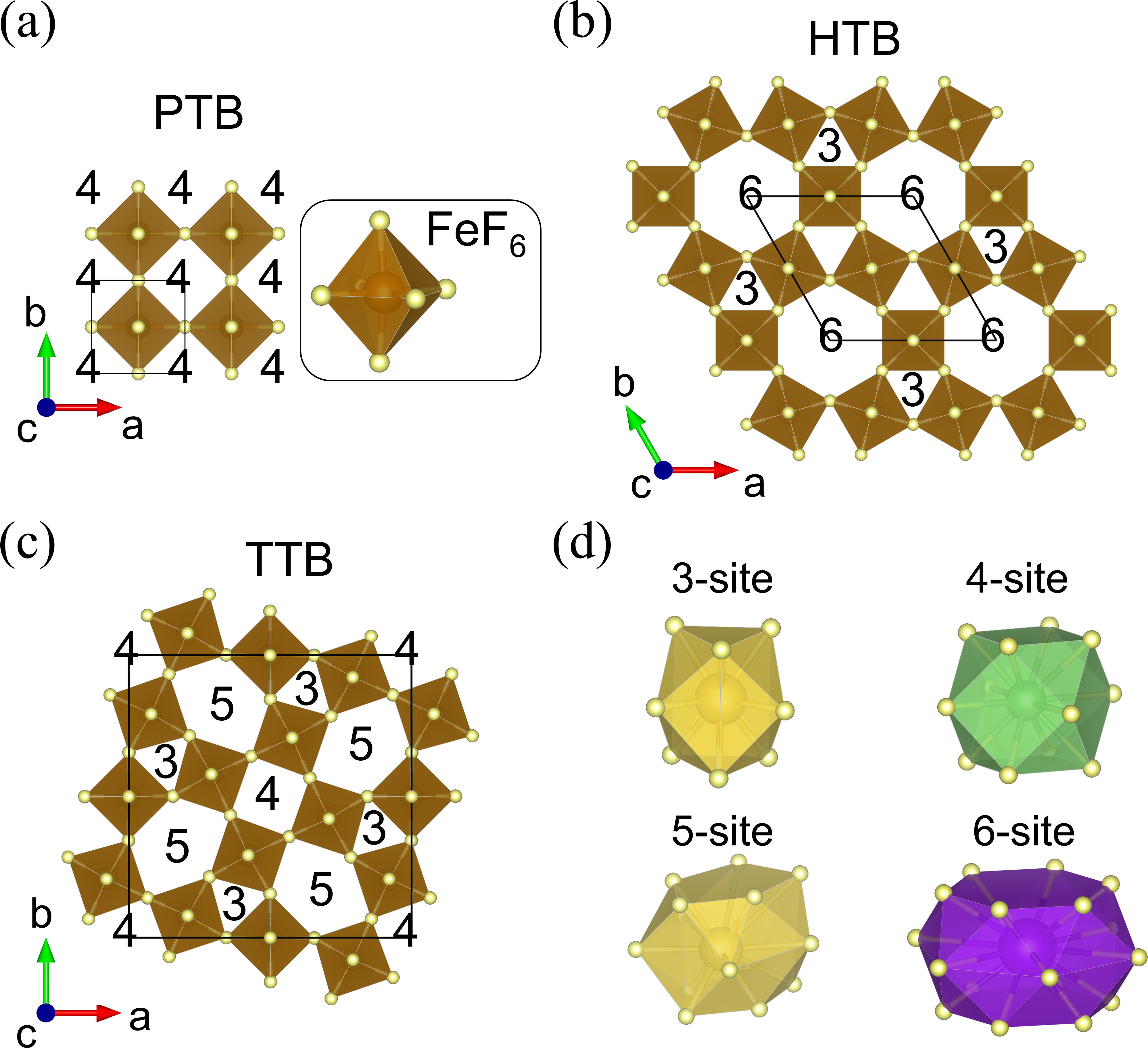}
    \caption{The (a) PTB, (b) HTB, and (c) TTB structures viewed along the c-axis with unoccupied intercalation sites ($A$ sites), indicated by numbers. Solid lines mark the unit cells. (d) Fluorine (light yellow spheres) polyhedra surrounding the intercalation sites.}
    \label{fig1:TB_strucs}
\end{figure}

Figures \ref{fig1:TB_strucs} (a), (b) and (c) show the three TB-type structures investigated in this work, projected along the $c$-axis. For each structure the unit cell corresponds to one layer of the region marked by the dashed lines. The possible intercalation sites, which form channels along the $c$-axis, are indicated by numbers. They correspond to the number of fluorine neighbors in the $ab$-plane and increase with increasing associated volume. The centers of those sites are surrounded by F$\mathrm{^-}$ anions that form the polyhedra $A$F$_j$, where $j$ inidicates the number of F neighbors of each site type and is equal to 9, 12, 15 and 18 for the 3-, 4-, 5- and 6-sites, respectively. The fluorine polyhedra are shown in Fig.\  \ref{fig1:TB_strucs} (d). For the TTB structure another notation is used in literature, where the {3-}, 4- and 5-sites are labeled C, A1 and A2 (cf.\ e.g., \cite{TTB_nomenclature}). However, for better comparability of structures and association of a relative volume with a number, we employ the numerical nomenclature in this paper.

The unit cells of the pristine crystal structures of PTB, HTB and TTB, i.e.\ before any distortions due to forces and stress minimization, have symmetries corresponding to the space groups $Pm\overline{3}m$ (221), $P6/mmm$ (191) and $P4/mbm$ (127), respectively. The initial dimensions (in \AA) of the unit cells along $a$, $b$ and $c$ are \nolinebreak{[4.2, 4.2, 4.2]} for PTB, \nolinebreak{[7.5, 7.5, 3.9]} for HTB and \nolinebreak{[12.8, 12.8, 4.0]} for TTB. Table \ref{tab:wyckoff_positions} lists the Wyckoff positions and representative coordinates of the lattice sites in the three structures.

\begin{ruledtabular}
\begin{table}
\centering
\begin{tabular}{ llccccc} 
 & Atom & Site & x & y & z \\
\hline
PTB     &$A$ (4-site)  & 1a & 0 & 0 & 0\\
        &Fe  & 1b & 1/2 & 1/2 & 1/2\\ 
        &F & 3c & 0 & 1/2 & 1/2\\
    \hline
HTB     &$A$ (6-site)  & 1a & 0 & 0 & 0\\
        &$A$ (3-site)  & 2c & 1/3 & 2/3 & 0\\ 
        &Fe  & 3g & 1/2 & 0 & 1/2\\
        &F1 & 3f & 1/2 & 1/2 & 0\\
        &F2 & 6l & 0.212 & 0.423 & 0\\
    \hline
TTB     &$A$ (3-site) & 4g & 0.125&   0.625&   0\\ 
        &$A$ (4-site) & 2a & 0 & 0 & 0\\  
        &$A$ (5-site) & 4h & 0.173 &  0.327 & 0\\     
        &Fe1  & 2c & 1/2 & 0 & 1/2\\ 
        &Fe2  & 8j & 0.214 & 0.075 & 1/2\\ 
        &F1   & 8j & 0.143& -0.068 & 1/2\\
        &F2   & 4h & 0.278& 0.222& 1/2\\ 
        &F3   & 8j & 0.344& -0.008& 1/2\\ 
        &F4   & 2d & 0.5 &   0  & 0 \\
        &F5   & 8i & 0.208 & 0.077 & 0\\                    
\end{tabular}
\caption{Wyckoff positions and representative fractional coordinates of the lattice sites for the three TB structure unit cells. The atomic coordinates for the fluorine atoms labelled F2 in HTB were adapted from Ref.\ \cite{Leblanc_1983_HTB_coords} and the atomic coordinates for TTB from Ref.\ \cite{Reisinger_2011_TB_dist}, except for the $A$ (3-site) which we placed in the center of the 3-sites.}
\label{tab:wyckoff_positions}
\end{table}
\end{ruledtabular}

\subsection{Computational details}\label{sec:methods_compdetails}

In order to allow for different magnetic orderings of the iron ions in PTB, we have used a 2$\times$2$\times$2 supercell of the unit cell depicted in Fig.\ \ref{fig1:TB_strucs} (a). For HTB and TTB we used 1$\times$1$\times$2 supercells of the unit cells in Fig.\ \ref{fig1:TB_strucs} (b) and (c), to allow for magnetic ordering along the $c$-axis. These supercells correspond to 8, 6 and 20 formula units for PTB, HTB and TTB, respectively.

We calculate energy-volume (EV) curves \cite{Silvi_EOS,TYUTEREV_EOS_2006} to determine equilibrium supercell volumes $V_0$ and energies $E_0$. This procedure, instead of typical stress minimization routines implemented in most atomistic simulation codes, was chosen to avoid systematic errors arising from Pulay stresses, which originate from the incompleteness of the plane-waves basis. We determine $V_0$ and $E_0$ by fitting the Murnaghan equation of state (EOS) \cite{Murnaghan_EOS_1944} to a set of ground state energies for at least seven volumes. For each volume, the cell shape and atomic coordinates were optimized. The local (polyhedron) volumes are extracted from the structures optimized at the equilibrium volume.

The DFT calculations were carried out with the \textit{Vienna ab initio simulation package} (VASP) \cite{vasp_general}. We used the Perdew-Burke-Ernzerhof generalized gradient approximation (PBE-GGA)\cite{perdew_generalized_1996} as the exchange-correlation functional. Interactions between valence electrons and ionic cores are treated with the projector-augmented waves (PAW) method \cite{vasp_paw}. As valence states we took the \textit{1s} and \textit{2s} orbitals for Li, the \textit{2s}, \textit{2p} and \textit{3s} orbitals for Na, the \textit{3s}, \textit{3p} and \textit{4s} orbitals for K, the \textit{3d} and \textit{4s} orbitals for Fe, and the \textit{2s} and \textit{2p} orbitals for F.

A Hubbard-$U$ correction term was applied to the \textit{3d} orbitals of Fe to mitigate the self-interaction error occurring, similar to oxides, in fluorides of transition metals with strongly localized orbitals \cite{Anisimov_1991_DFTU,Tolba18_DFTU,Himmetoglu_2013_DFTUreview,Correa_2018_DFTU}. We followed the DFT+$U$ approach of Dudarev \textit{et al.} \cite{dudarev_electron-energy-loss_1998} and set $U\mathrm{_{eff}} = U - J = 5$ eV, in accordance with previous works concerned with iron-based fluorides of TB type structures \cite{Yamauchi_2010_DFTU,Burbano_2015_DFTU,Zhang_2018_DFTU}. $U$ denotes the on-site Coulomb term and $J$ the site exchange term.

All calculations were spin-polarized (collinear). For all the compounds analyzed, comparative calculations showed that the iron ions are present in the high-spin state (i.e.\ maximizing the number of singly occupied \textit{3d} orbitals resulting in the highest possible magnetic moments per iron ion for the given number of electrons). The distribution of Fe$\mathrm{^{2+}}$ and Fe$\mathrm{^{3+}}$ ions was not specified prior to relaxation. After the relaxation of the ionic and electronic degrees of freedom, Fe ions were identified as Fe$\mathrm{^{2+}}$ or Fe$\mathrm{^{3+}}$ from their final magnetic moments.

We set the energy cutoff for the plane-waves basis to 700 eV and used a 4$\times$4$\times$4 (for PTB and HTB) and a 3$\times$3$\times$4 (for TTB) Monkhorst-Pack \cite{monkhorst} \textit{k}-point mesh ($>$~25 \textit{k}-points/\AA$\mathrm{^{-1}}$ for all structures and lattice directions) with a Gaussian smearing of 0.05 eV \cite{Fu_1983_smearing,Elsaesser_1994_smearing} for the Brillouin-zone integrations. We converged the total energies to 5$\times$10$^{-6}$~eV in the electronic self-consistency loop and the interatomic forces to 5$\times$10$^{-5}$~ eV/{\AA} in the ionic relaxation loop.


\section{Results} \label{sec:results}

\subsection{Magnetic ordering of the Fe ions}\label{res:results_magnetic}

\begin{figure}
    \centering
    \includegraphics[width=0.99\columnwidth]{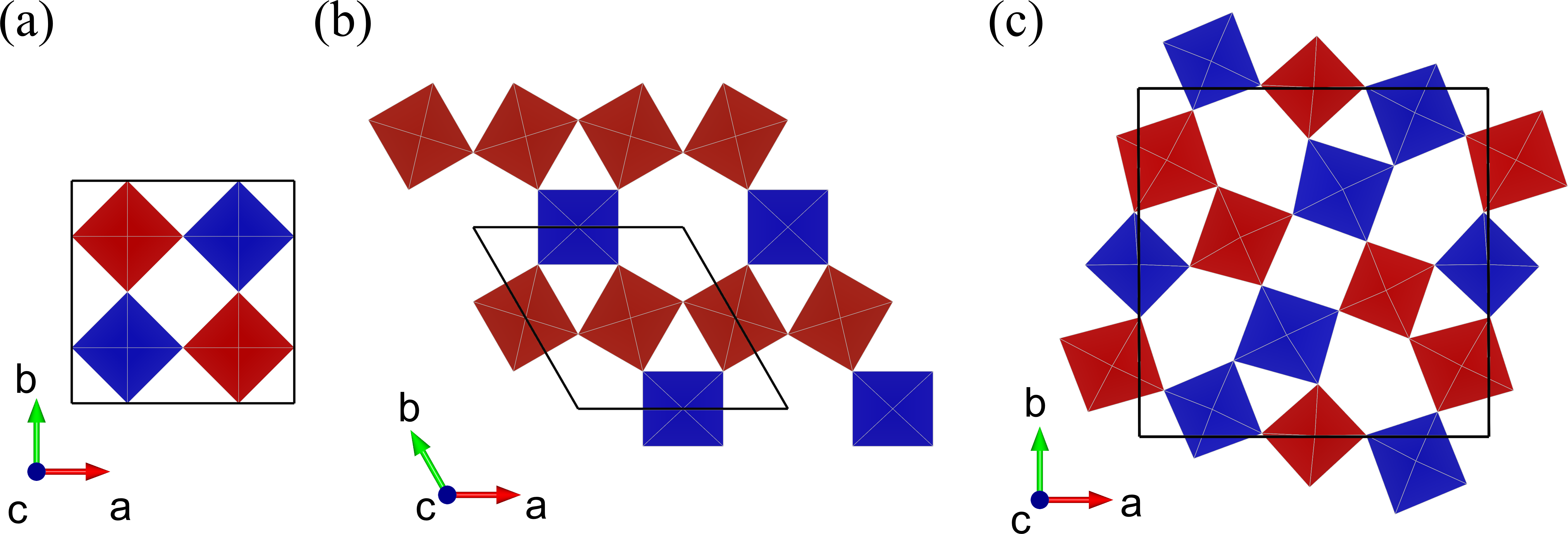}
    \caption{Arrangement of magnetic moments of the iron ions in TB-type structures PTB (a), TTB (b) and HTB (c). Blue and red octahedra correspond to central Fe ions with spin-up and spin-down, respectively. Solid lines mark the outline of the supercells in the $ab$-plane. Only the layer at $z\approx 0.25$ is shown, for the layer at $z\approx 0.75$ the colors are inverted.}
    \label{fig2:mag_arrangments}
\end{figure}

Different magnetic arrangements can lead to different equilibrium volumes $V_0$, and thus to different relative volume changes 
\begin{equation}
     \Delta V_0 = \frac{V_0^{\mathrm{final}}-V_0\mathrm{^{initial}}}{V_0\mathrm{^{initial}}} \times 100\%.
\end{equation}
Here, $V_0\mathrm{^{initial}}$ and $V_0\mathrm{^{final}}$ are the supercell volumes at lower and higher $x$ concentrations, respectively, and thus $\Delta V_0$ describes the volume change associated with a change in concentration. We first determined the magnetic ground states of the structures. For PTB, we investigated the ferromagnetic (FM) and four antiferromagnetic (AFM) arrangements of the iron magnetic moments (a description of the arrangements can be found in Ref.\ \cite{george_fundamentals_2020_perosvkites}). Table \ref{tab:PTB_magnetic} lists the equilibrium energies and volumes for the magnetic arrangements in the empty (PTB FeF$\mathrm{_3}$) and filled structures, here given for Na on the $A$ site (PTB NaFeF$\mathrm{_3}$). It can be seen that although the differences in energy are small, the differences in $\Delta V_0$ amount up to 1.4 \%. This value is in the range of volume changes obtained in this work, highlighting the importance of describing the proper magnetic ground states. For the PTB structure the ground state corresponds to the G-type AFM ordering, in which the magnetic moments change sign for neighbors along all three spatial directions. This arrangement is visualized in Fig.\ \ref{fig2:mag_arrangments} (a). Exemplary calculations showed that the energetic order of the states does not depend on the type of $A$ cations.

\begin{ruledtabular}
\begin{table}
\centering
\begin{tabular}{lccccc} 
 &\multicolumn{2}{c}{PTB FeF$\mathrm{_3}$} & \multicolumn{2}{c}{PTB NaFeF$\mathrm{_3}$} & \\
 & $V_0$ [\AA$\mathrm{^3}$] & $E_0$ [eV] & $V_0$ [\AA$\mathrm{^3}$]& $E_0$ [eV]&$\Delta V_0$[\%]\\
\hline
A-AFM & 58.47 & 0.095& 69.73 & 0.018 & 19.26\\
C-AFM & 58.30 & 0.046& 68.90 & 0.008 & 18.17\\
E-AFM & 58.31 & 0.047& 69.64 & 0.009 & 19.44\\
G-AFM & 58.14 & 0.0  & 69.55 & 0.0   & 19.61\\
FM    & 58.64 & 0.146& 69.84 & 0.03  & 19.10  \\

\end{tabular}
\caption{Equilibrium volumes and energies for the empty and sodiated PTB structure for different arrangements of the magnetic moments of the Fe ions. All quantities are given per formula unit. The energies are referenced to the lowest energy configuration, the G-type AFM ordering.}
\label{tab:PTB_magnetic}
\end{table}
\end{ruledtabular}

We have calculated the energy of the FM ordering and of all the 20 possible AFM arrangments, in the empty and filled HTB supercells. The ground state arrangement is depicted in Fig. \ref{fig2:mag_arrangments} (b). It exhibits alternating signs for the Fe magnetic moments along the $c$ axis, and the moments of iron ions with the same $a$ and $c$ coordinates have the same direction.

For TTB, the 20 iron ions in the supercell result in too many combinatorial configurations to treat them all in the scope of this work. We therefore took the experimentally measured arrangement of the iron magnetic moments reported by Mezzadri \textit{et al.} for K$\mathrm{_{0.6}}$FeF$\mathrm{_{3}}$ \cite{Mezzadri_2011_TTB} as orientation. Differently to the experiments, we considered collinear arrangements. We set up the initial magnetic moments as depicted in Fig. \ref{fig2:mag_arrangments} (c). The arrangement of the moments did not change after force and stress minimization, and for K$\mathrm{_{0.6}}$FeF$\mathrm{_{3}}$ we obtained a ferrimagnetic ordering, in agreement with Mezzadri \textit{et al.}. The net magnetization amounts to 2 $\mu\mathrm{_B}$ per supercell. Compared to a FM arrangement for the same compound, this arrangement lowers the energy by~0.75 eV and we therefore chose this initial arrangement for all TTB calculations. All three TB structures have in common that they have alternating directions of the magnetic moments along the $c$-axis, underlining the importance of the choice of the supercells used in this work to describe the proper magnetic ground states.

\subsection{TTB Na$_x$FeF$\mathrm{_3}$}\label{sec:result_naxfef3}

We start our investigation with the TTB Na$_x$FeF$\mathrm{_3}$ and compare the results with the experimental data of Han \textit{et al} \cite{Han_2016_ZS_TB}. 
The 20 possible intercalation sites in the TTB supercell would result in too many possible inequivalent configurations to be treated within the scope of this work to determine the energetic order of occupation of individual sites. Therefore, we use the approach that first all sites of the site type with the lowest formation energy $E\mathrm{_{form}^{Na^+}}$ are filled, then the sites with the next lowest formation energy in the now partially occupied structure, and finally the sites of the remaining site type. The formation energy per $A$ cation is defined as follows:

\begin{equation}
     E_{\mathrm{form}}^A= \frac{E_{\mathrm{final}}^{A_x\mathrm{FeF_3}}-E_{\mathrm{initial}}^{A_x\mathrm{FeF_3}}- n E_{\mathrm{bulk}}^A}{n}.
\end{equation}

\begin{figure}
    \centering
    \includegraphics[width=0.8\columnwidth]{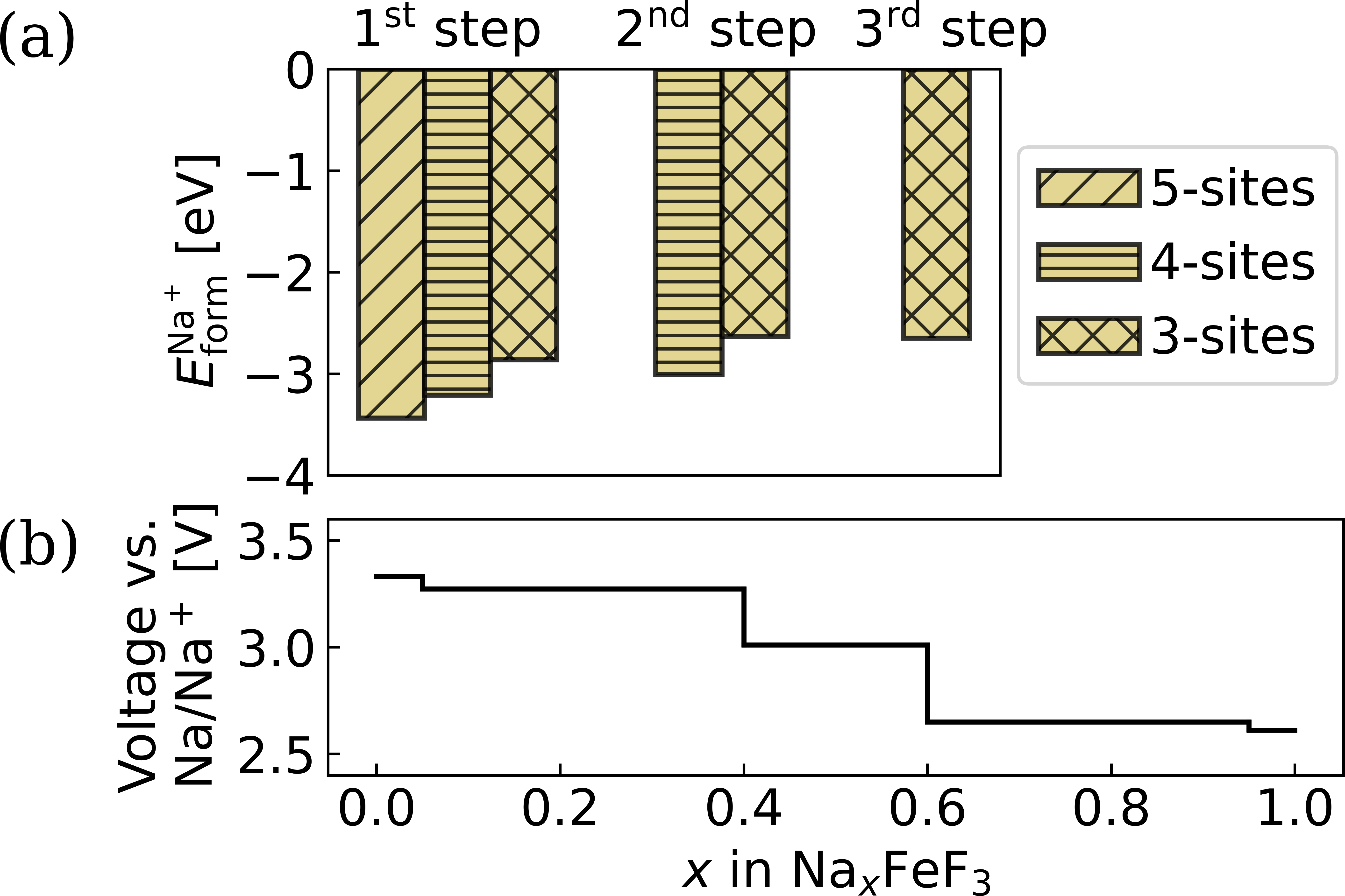}
    \caption{(a) Formation energies per Na cation, $E\mathrm{_{form}^{Na^+}}$, for the three site types in the TTB structure. In each step the sites with lowest formation energy are occupied. (b) Voltage w.r.t.\ Na/Na$\mathrm{^+}$ for intercalation of Na cations into the TTB structure. The concentrations 0.05, 0.4, 0.6, 0.95 and 1 correspond to the occupation of one 5-site, all 5-sites, all 4-sites, all but one 3-site, and all 3-sites, respectively.}
    \label{fig3:naxfef_energies}
\end{figure}

Here, $E_{\mathrm{initial}}^{A_x\mathrm{FeF_3}}$ and $E_{\mathrm{final}}^{A_x\mathrm{FeF_3}}$ are the energies at lower and higher $x$ concentrations, respectively, and $n$ is the number of intercalated cations, related to $\Delta x$ by \nolinebreak{$\Delta n=x \times$formula units}. The formation energy per Na cation is shown in Fig.\ \ref{fig3:naxfef_energies} (a) for the three intercalation steps. In the first step the 5-sites are occupied. The 4-sites have the lower formation energy in the 5-site occupied structure and are therefore occupied next, followed by the 3-sites. 

Figure \ref{fig3:naxfef_energies} (b) shows the voltage profile for this stepwise occupation, w.r.t.\ Na/Na$\mathrm{^+}$, i.e. assuming a metallic sodium anode. The first (short) horizontal line corresponds to the occupation of one single 5-site and the last (short) horizontal line to all sites being occupied, except one single 3-site. The very small differences to the average values of the occupation of all resepective sites demonstrate the validity of the above described procedure of stepwise site-type occupation. The average voltage values are in good agreement with the values of Han et al. \cite{Han_2016_ZS_TB} who reported voltages between 2.5 and 4 V in the concentration range $x$=0 to $x$=0.6. In agreement with the geometric arguments given in that paper that the 3-sites are too small for suitable Na$\mathrm{^+}$ storage, the occupation of this site-type showed the lowest formation energy.

\begin{figure}
    \centering
    \includegraphics[width=1\columnwidth]{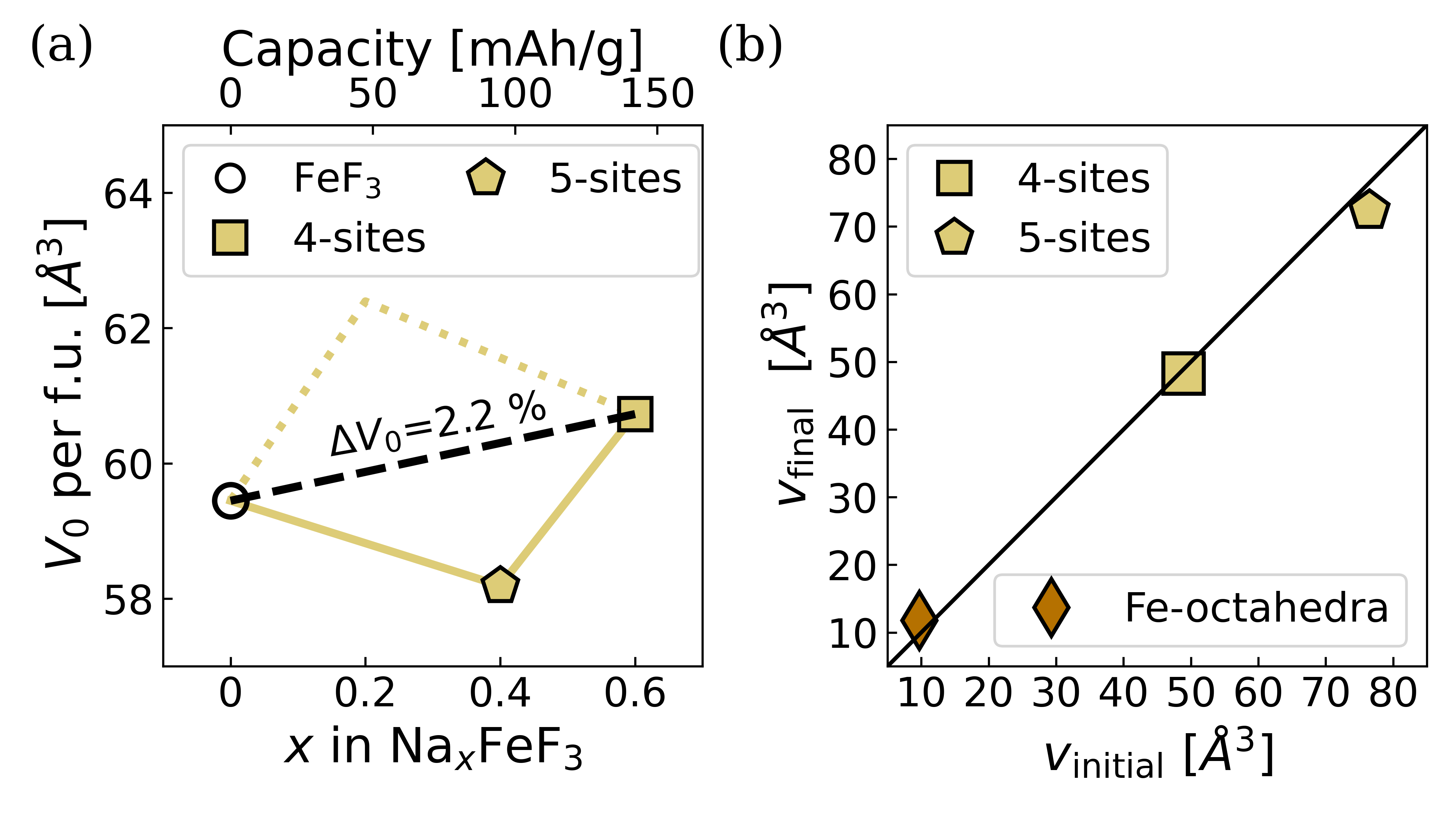}
    \caption{(a) Unit cell volumes for Na$_x$FeF$\mathrm{_3}$ in the TTB structure. The dotted line shows the energetically less favorable course if the 4-sites were occupied first and then the 5-sites. (b) Change in local volumes of the Na$_x$FeF$\mathrm{_3}$ structure. For the FeF$\mathrm{_6}$ octahedra \textit{initial} refers to the oxidation state Fe$\mathrm{^{3+}}$ and \textit{final} to Fe$\mathrm{^{2+}}$. For the intercalation sites, \textit{inital} and \textit{final} refer to the average volume per site type of the F polyhedra surrounding the sites before and after occucpation by Na cations, respectively.}
    \label{fig4:naxfef_volumes}
\end{figure}

The global volumes (represented by uppercase $V$'s) up to a concentration of $x$=0.6 are shown in Fig.\  \ref{fig4:naxfef_volumes} (a). \textit{Global} in this context means the volume of the supercell and thus stands for the expansion of a hypothetical perfect single crystal. The occupation of the 5-sites initially leads to a reduction in volume, while the subsequent occupation of the 4-sites leads to an expansion in volume. Overall, this results in a global volume change of 2.2\% for a capacity comparable to that measured by Han et al. ($\sim$150 mAh/g). This result is in good agreement with the reported near-ZS behavior for this compound. In reality the structure is rarely a perfect single crystal, and thermal and kinetic effects play a role. If the 4-sites were occupied first, the material would expand to a volume of 62.4 \AA~per unit cell, as indicated in Fig.\  \ref{fig4:naxfef_volumes} (a) with dotted lines. Occupying the 3-sites as first sites would lead to an even larger volume expansion of 67.6 \AA~per unit cell, but this is associated with a formation energy of -2.7 eV/atom as opposed to -3.3 eV/atom for the 5-sites. Assuming a Boltzmann distribution, at a temperature $T$ the probability of finding a system in a state with energy $E$ is proportional to $e^{(-E/k\mathrm{_{B}}T)}$. At room temperature $k\mathrm{_{B}}T$ is about 0.025 eV, which is an order of magnitude smaller than the differences in formation energies. Therefore, we will take the lowest states as a good first approximation to the behavior of the material, but note that in reality a more mixed occupation can be expected for intermediate concentrations, e.g., some occupation of 4-sites in TTB Na$_x$FeF$\mathrm{_3}$ before all 5-sites are occupied, which would lead to a volume change curve lying between the two limit cases outlined in  Fig.\  \ref{fig4:naxfef_volumes} (a).

To elucidate the mechanisms responsible for this small global volume change, we look at local volumes (represented throughout the text by lowercase $v$'s) within the structure. In Figure \ref{fig4:naxfef_volumes} (b), the volumes of FeF$\mathrm{_6}$ octahedra are shown for the iron ions that change their oxidation state during Na incorporation. To achieve charge neutrality, a number of iron atoms corresponding to the number of Na$\mathrm{^+}$ cations must change their oxidation state from 3+ to 2+. The change in oxidation state leads to an increase of the ionic radius and thus to a stretching of the Fe-F bonds, which in turn increases the octahedron volumes surrounding the iron ions. This local volume expansion alone would lead to a global volume expansion. Therefore, in order to achieve small global volume changes, there must be a volume-reducing effect that compensates for this expansion.

During the intercalation of sodium into the sites, the volumes of the $A$F$_j$ polyhedra change. These changes are displayed in Figure \ref{fig4:naxfef_volumes} (b), where we show here and in the following the average volume per site type for better visibility (the typical variance between sites of the same site type is on the order of tenths of \AA$\mathrm{^3}$). \textit{Unoccupied} indicates the local volume of the site-type in the structure from the immediately preceding concentration, i.e. the global structure may be partially occupied. For both, 4-sites and 5-sites, the volume is smaller after occupation by Na than before. This can be explained by an electrostatic argument: There is an attraction between the positively charged Na$\mathrm{^+}$ cations and the negatively charged F$\mathrm{^-}$ anions, and the intercalated sodium cation reduces the mutual repulsion of neighboring fluorine anions. This reduction in the local volume can partially compensate for the expansion of the iron octahedra, resulting in a small global volume expansion of the unit cell. For the occupation of the 5-sites this effect is more pronounced and therefore even leads to a global volume reduction, while the occupation of the 4-sites only results in a low local volume contraction and therefore does not completely compensate the expanding iron octahedra.

\subsection{Intercalation of Li$\mathrm{^+}$, Na$\mathrm{^+}$, and K$\mathrm{^+}$ in TB structures}\label{sec:result_systematic}

\begin{figure}
    \centering
    \includegraphics[width=0.99\columnwidth]{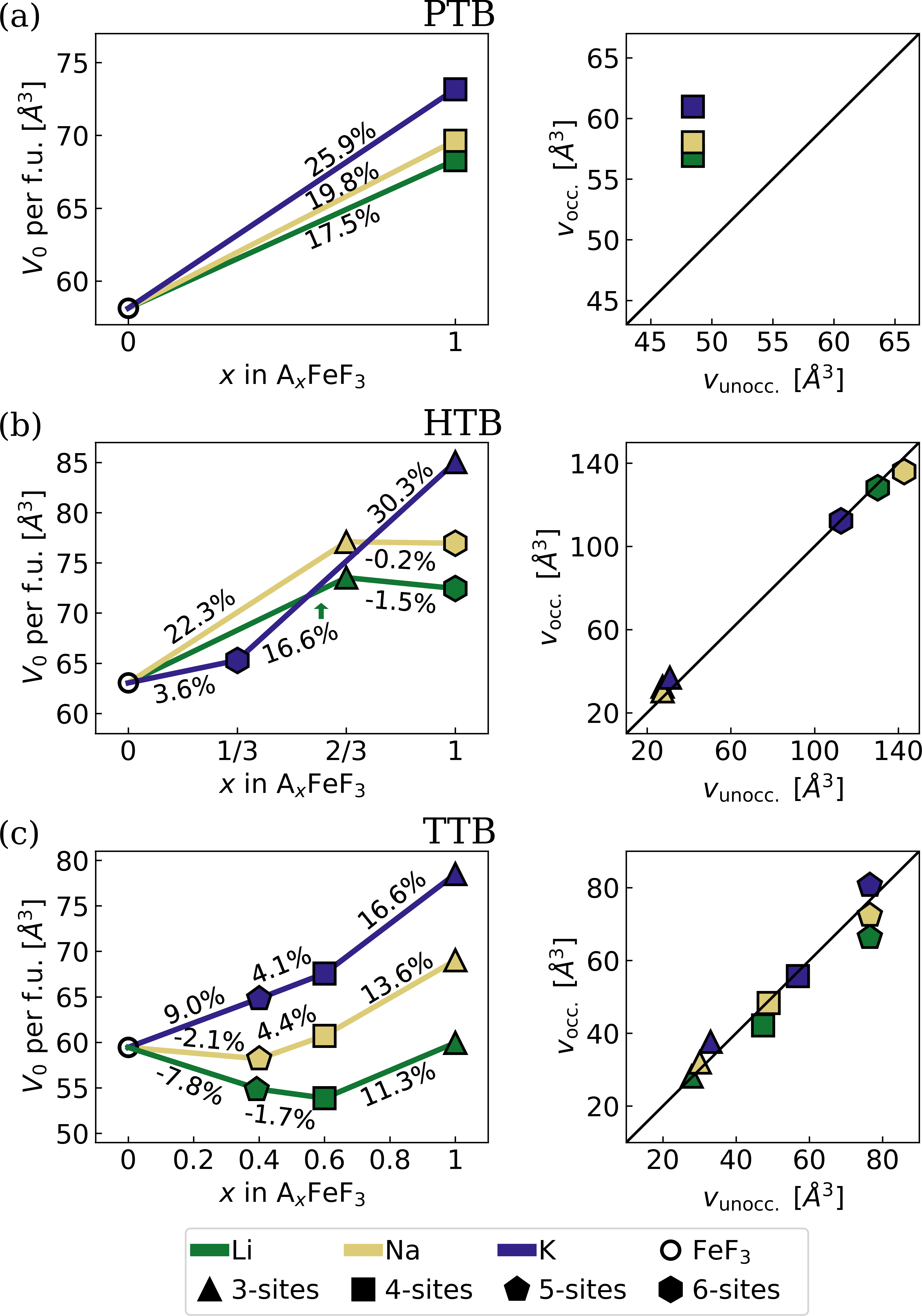}
    \caption{Global and local volumes of $A_x$FeF$\mathrm{_3}$ in the structures (a) PTB, (b) HTB and (c) TTB. Left: Unit cell volumes at different concentrations of the intercalated cations Li$\mathrm{^+}$, Na$\mathrm{^+}$ and K$\mathrm{^+}$. Symbols indicate the site type being occupied in the step leading to the corresponding concentration $x$. The values for $\Delta V_0$ are given as percentage for each intercalation step.  Right: Occupied vs. unoccupied average volumes per site type of the F polyhedra surrounding the different intercalation sites.}
    \label{fig5:systematic_volumes}
\end{figure}

Motivated by the consistency of our results with the experiment, we extended the investigation to the intercalation of Li$\mathrm{^+}$, Na$\mathrm{^+}$ and K$\mathrm{^+}$ in all three tungsten bronze structures PTB, HTB and TTB. In all cases, an expansion of the fluorine octahedra around the iron ions of decreased oxidation states is observed as described above. Hence, to achieve a low global volume expansion, a similar compensation mechanism as for TTB Na$_x$FeF$\mathrm{_3}$ must take place. In Figure \ref{fig5:systematic_volumes}, right panels, we show the change in volume, that is spanned by the F$\mathrm{^-}$ anions surrounding the $A$ sites, upon intercalation of $A$ cations. The left panels in Fig.\ \ref{fig5:systematic_volumes} show the global volumes of the three structures in the concentration ranges between $x$=0 and $x$=1. The order of occupancy of the sites was carried out stepwise according to the site types as described in Sec.\ \ref{sec:result_naxfef3}.

For the PTB structure (Fig.\ \ref{fig5:systematic_volumes} (a)), the global volumes (left panel) increase between 17\% and 26\%, depending on the type of intercalated $A$ cation. These global volume changes can be easily related to the 4-site volume changes (right panel). In all cases, the volume of the site after occupation by $A$ cations is larger than the volume of the empty site. There is also a clear trend towards larger local and global volumes with larger intercalated cation. The ionic radius increases from Li to Na and to K. Since there is no local volume reduction, there is no compensation mechanism and therefore a large global volume increase.

In the HTB structure (Fig.\ \ref{fig5:systematic_volumes} (b)), different types of sites are occupied first, depending on the type of the $A$ cation (left panel). For the two smaller cations Li and Na the 3-sites are occupied first, in the case of K the 6-sites. For all cations, the occupation of the smaller 3-sites leads to a larger global increase in volume than the occupation of the 6-sites. Again, this global volume phenomenon can be well related to the local changes shown in the right panel of Fig.\ \ref{fig5:systematic_volumes} (b): for the 3-sites, the volume increases slightly, whereas the occupation of the 6-sites leads to a local volume decrease for all $A$ cations. In the case of Li, this decrease in volume is most pronounced, which is reflected in the decrease of the global volume by 1.5\% from $x$=2/3 to $x$=1.

In the TTB structure (Fig.\ \ref{fig5:systematic_volumes} (c)), the energetic order of occupation of the different site types is the same for all $A$ cations: The 5-sites are occupied first, then the 4-sites, and finally the 3-sites. In the case of $A$=Li, the filling of both, the 5- and the 4-sites, leads to a global volume reduction, whereas the filling of the 3-sites leads to an expansion of the cell volume. Again, this correlates with the local $A$F$_j$ volumes: The reduction is largest by filling the  5-sites, followed by the 4-sites. The occupation of the 3-sites leads to a local volume reduction so small that, in combination with the expanding FeF$\mathrm{_6}$ octahedra, a global volume expansion occurs. In the case of $A$=Li, there is a change in cell volume of only 0.8\% over the entire concentration range from $x$=0 to $x$=1. The occupation of the 5- and 4-sites by Na has already been described in the Sec.~\ref{sec:result_naxfef3}. The occupation of the {3-sites} by Na leads to a small volume increase of the NaF$\mathrm{_9}$ polyhedra and thus to a considerable global increase of the cell volume of about 13\%. For $A$=K, the occupation of each site type leads to a global increase in volume. Only in the case of the 4-sites there is a minimal local volume decrease, which however cannot compensate for the expanding FeF$\mathrm{_6}$ octahedra for this site type.

\section{Discussion}\label{sec:discussion}

\subsection{Suitability of TB structures as cathode materials}\label{disc:TB_as_cathode}

The described systematic investigation of structures, site types and intercalated ions leads us to a discussion of the suitability of the investigated compunds with TB structures as ZS cathode materials. In the perovskite structure there is no mechanism that compensates for the expansion of the FeF$\mathrm{_6}$ octahedra, and hence the PTB structure experiences a large increase in volume upon intercalation of all types of the considered $A$ cations. According to our calculations, the completely empty structure is therefore not a promising ZS cathode material. However, the situation may be different if the structure is expanded by "filler cations", as descriped by Martin et al. \cite{Martin_2019_PTB_K_cathode}, where K$\mathrm{^+}$ and NH$\mathrm{_4^+}$ were used during synthesis and the material was subsequently cycled vs.\ lithium. Cao and coworkers also employed K$\mathrm{^+}$ cations to pre-expand the framework for later Na$\mathrm{^+}$ cycling \cite{Cao_2017_PTB_cathode}. Although no ZS behavior was reported in either of those works, the enlarged sites may allow for a stronger compensation mechanism leading to less drastic volume changes compared to the values reported here.

In the HTB structure, there is a compensation mechanism for the occupation of 6-sites, which leads to global volume changes of -1.5 \% and -0.2 \% for $A$=Li and $A$=Na, respectively. However, there is only one 6-site per unit cell, which corresponds to $\Delta x$ =1/3 per formula unit. This means that there is a global volume increase below $x$=2/3 (occupation of the 3-sites) and the small volume changes occur above $x$=2/3. For the compound FeF$\mathrm{_3}$-0.33H$\mathrm{_2}$O, Li et al. reported a volume expansion of $\sim$3.9\% during the intercalation of $\Delta x$=0.66 Li and found indications that up to two Li$\mathrm{^+}$ cations occupied one 6-site \cite{Li_2011_Li_cathode}. We have not considered the possibility of double occupancy in this work, but we expect an expansion of the local $A$F$_j$ volume (as opposed to contraction with single occupancy reported here) and thus a global expansion of the structure. Consequently, only low capacities with simultaneous ZS behavior are achievable in the HTB structure.

For TTB Na$_x$FeF$\mathrm{_3}$ the ZS behavior was already measured experimentally and explained above by our atomistic simulations. In the case of K$\mathrm{^+}$, there is no site type that provides a sufficient compensation mechanism to maintain small global volume changes. For Li, however, both occupations of the 4-sites and of the 5-sites lead to a global volume reduction, while the occupation of the 3-sites expands the cell volume. This results in a small global volume change of 0.8 \% for the entire concentration range. For intermediate concentration changes, we calculated larger volume changes, far beyond ZS behavior, when the occupation occurs strictly stepwise by site-type (e.g., $\Delta V_0$=-7.8\% from \textit{x}=0 to \textit{x}=0.4 for the occupation of the 5-sites, see Fig.\ \ref{fig5:systematic_volumes} (c), left). However, as stated above, it is reasonable to expect a more mixed occupation, and hence less pronounced global volume changes at intermediate concentrations (see also Fig.\ \ref{fig4:naxfef_volumes} (a)). Thus, TTB Li$_x$FeF$\mathrm{_3}$ represents a promising candidate material for ZS applications, with a theoretical capacity of 223 mAh/g and an average voltage of 3.23 V. For smaller cations, which show global volume contractions upon intercalation of certain site-types, a ZS mechanism can be achieved due to mixed occupation and hence the iron-based fluorides TB structures are not suitable ZS materials for the large K$\mathrm{^+}$ cations and for Na$\mathrm{^+}$ cations only for limited concentration ranges.

\subsection{ZS design criteria}\label{disc:quali_ZS}

In Section \ref{sec:result_systematic} we described what affects the local volume changes: (i) larger cations lead to larger local volumes (see, e.g., Fig.\ \ref{fig5:systematic_volumes} (a), right) and (ii) sites with too small local volumes may prevent a volume reduction, while larger local volumes may allow an appropriate compensation mechanism (see, e.g., Fig.\ \ref{fig5:systematic_volumes} (b), right). However, these two criteria alone do not allow a qualitative classification of whether the local volume decreases, remains the same, or increases upon occupation. For example, the 4-sites in PTB and TTB show very different behavior. In the former structure there is a strong volume expansion, while in the latter case the volumes are nearly constant before and after the incorporation of the $A$ cations. In order to obtain a structure descriptor that can be used to predict the qualitative prerequisite for a compensation mechanism (local reduction of the $A$F$_j$ volume), it is thus also necessary to describe the properties of the structures itself.

\begin{figure}
    \centering
    \includegraphics[width=0.8\columnwidth]{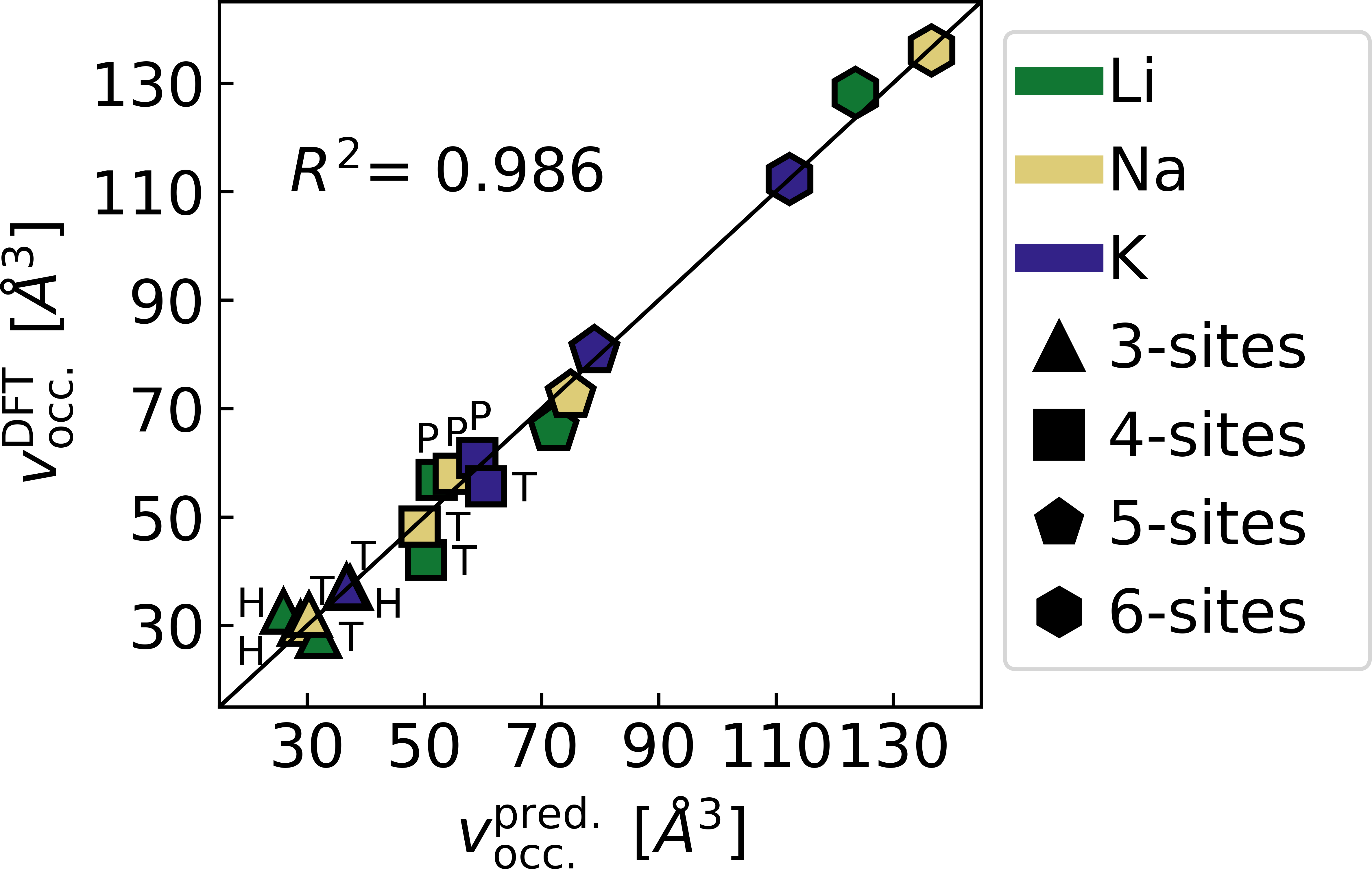}
    \caption{Local volumes after occupation by $A$ cations from DFT calculations versus predicted values. Colors and markers represent the intercalated cation and site type, respectively. For site types that occur in several structures, the host structure (P: PTB, H: HTB, T: TTB) is labeled besides the data points.}
    \label{fig6:design_criteria}
\end{figure}

We used various features (e.g., bond lengths, symmetry and density features) as input for a number of structure descriptors which are implemented in the Python package \textit{matminer} \cite{Matminer_Ward_2018}. Each descriptor takes a dataset of structural features as input and returns a numerical digit describing some aspect of the global structure. These values can then be used in numerical processing, such as analysis or building predictive models. Each of these descriptors was tested separately together with two parameters, the ionic radius of the $A$ cation $r\mathrm{_{ionic}^\textit{A}}$ and the unoccupied site volume $v\mathrm{_{unocc.}}$, in a linear regression model to find that structure descriptor, which results in the best prediction (i.e., lowest residual sum of squares, $R^2$) of the set of local volumes compared to the DFT data. Applying the Python package \textit{sklearn} for this task \cite{scikit-learn}, the descriptor MAD (mean absolute deviation of relative bond length) was identified as the optimal choice, resulting in a $R^2$ of 0.986. The linear equation of the predictor for standardized input data (removing the mean value and scaling to unit variance) was obtained as follows:
\begin{equation}\label{eq:vol_predictor}
    v\mathrm{_{occ.}^{pred.}} =  2.85*r\mathrm{^\textit{A}_{ionic}} + 34.68*v\mathrm{_{unocc.}} -3.92*\mathrm{MAD}.
\end{equation}
Figure \ref{fig6:design_criteria} shows the values calculated with this predictor compared to the true values from our DFT calculations.

The signs of the coefficients of each feature can be interpreted as follows: A larger ionic radius means that the bonds to the surrounding fluorine anions are larger and therefore the volume will be larger for larger cations. Larger initial volumes obviously lead to larger occupied volumes (e.g.\ the 3-sites are always smaller than the 6-sites, before or after occupation), which corresponds to a positive correlation. If both, the site type and the type of intercalated cation, are the same (e.g., occupation of the 4-sites in PTB and TTB), the "flexibility" of the structure determines whether large volume changes can occur. This is described here by MAD, a measure for the structural heterogeneity: higher values correspond to structures with higher variance in bond lengths. For PTB this variation is smaller than for TTB, therefore the predicted volumes for the same site type and the same $A$ cation are lower in the latter case (negative sign for MAD in Eq.\ (\ref{eq:vol_predictor})). From the magnitude of the coefficients, it can be seen that the unoccupied volume has the greatest influence on the occupied volume, which can be interpreted to mean that the volumes are modified by intercalation, but not completely changed, suggesting a reversible mechanism.

As discussed above, for small global volume changes to happen, a sufficient compensation mechanism needs to take place. This is the case, when the local $A$F$_j$ volumes are reduced in a manner that compensates for the volume-expanding effect of the FeF$\mathrm{_6}$ octahedra. Therefore the ratio between $v\mathrm{_{occ.}^{pred}}$ and $v\mathrm{_{unocc.}}$ can guide towards suitable combinations of structure, unoccupied site volume, and $A$ cation, that result in small global volume changes. Values of the ratio below 1 indicate sites, for which the local volume reduction can either directly lead to ZS behavior of the structure ($v\mathrm{_{occ.}^{pred}}$/$v\mathrm{_{unocc.}}$=0.95 and 0.96 for occupation of 6-sites in HTB by Li and Na, respectively) or in combination with other sites to ZS behavior over larger concentration ranges ($v\mathrm{_{occ.}^{pred}}$/$v\mathrm{_{unocc.}}$=0.94 and 0.98 for occupation of 5-sites in TTB by Li and Na, respectively). Values of 1.1 and higher indicate structures, for which no suitable compensation mechanism can be expected. In this work, we have not discussed the case of structures, where different $A$ cations are present, e.g., where K cations serve as "filler cations", as discussed above. For example, for the TTB structure K$\mathrm{_{0.6}}$FeF$\mathrm{_{3}}$, where potassium occupies the 4- and 5-sites, the descriptor predicts a $v\mathrm{_{occ.}^{pred}}$/$v\mathrm{_{unocc.}}$ ratio of 0.903, for the occupation of the 3-sites by Li, indicating a possible compensating mechanism.

Although the proposed descriptor was only trained on iron-based fluorides with TB-type structures and is therefore not directly applicable to other transition metal-based fluorides/oxides, it shows that simple correlations can be developed for such systems and help to find possible ZS materials. We are confident that it represents a step towards quantitative design criteria for ZS behavior. In combination with the correlation found in Ref.\ \cite{Baumann_Col} between changes in the size of TM ions (Shannon radii) and the local volumes of the anion octahedra around them, a more complex but still computational inexpensive model can be envisioned that can predict possible ZS candidates in the subset of the \textit{3d} TM TB. Such models are not expected to be a complete replacement for DFT calculations, since, e.g., the determination of the magnetic ground state still relies on a calculation of the electronic structure but constitute a promising  first screening step.


\section{Conclusions} \label{sec:conclusion}

By employing atomistic simulations, we explained the experimentally reported ZS behavior in TTB Na$_x$FeF$\mathrm{_3}$ by means of a compensation mechanism. There are two opposite local volume changes during the intercalation of Na$\mathrm{^+}$ cations: Iron ions, which change their oxidation state due to global charge neutrality, lead to expanding FeF$\mathrm{_6}$ octahedra. This is opposed by local volume contractions of the sites occupied by sodium. The F$\mathrm{^-}$ polyhedra volumes surrounding the $A$ site decrease, due to the electrostatic attraction between the positive Na$\mathrm{^+}$ and the negative F$\mathrm{^-}$ anions.

The investigation of the volume change during intercalation was extended to three structures of the tungsten bronze family, PTB, HTB and TTB, and to three $A$ cations, Li$\mathrm{^+}$, Na$\mathrm{^+}$ and K$\mathrm{^+}$. The systematic study showed that, in contrast to the TTB structure, the completely empty PTB and HTB structures are not particularly suitable for ZS cathode applications. TTB Li$_x$FeF$\mathrm{_3}$ was identified and explained as a promising candidate, as ZS behavior was calculated for the concentration range from $x$=0 to $x$=1, corresponding to a high theoretical capacity of 223 mAh/g. To the best of our knowledge, the structure has not yet been experimentally studied for its suitability as a cathode material for Li ion batteries.

The systematic data allowed the development of a simple model to describe the local volume of $A$F$_j$ polyhedra after occupation by $A$. The ratio of $v\mathrm{_{occ.}}$/$v\mathrm{_{unocc.}}$ can be used as an indicator for structures with suitable compensation mechanisms. This represents a first step towards a model, that predicts the global volume change without expensive DFT calculations and thus an important step towards a faster screening of potential ZS alkali ion intercalation materials.


\acknowledgments
This work was funded by the German Research Foundation (DFG, Grant No.\ EL 155/29-1). The authors acknowledge support by the state of Baden-Württemberg through bwHPC and the German Research Foundation (DFG) through grant no INST 40/575-1 FUGG (JUSTUS 2 cluster). Crystallographic drawings were created with the software VESTA \cite{Momma_2011_VESTA}.\\

\bibliography{tb_paper_resubmission}

\end{document}